\documentclass[12pt,letterpaper]{article}
\usepackage[pdftex]{graphicx,color}
\usepackage{hyperref}
\usepackage{amsmath,amssymb}
\usepackage[dvips,letterpaper,text={6.5in,9in}]{geometry}
\usepackage{fancyhdr}
\usepackage{verbatim}

\usepackage{dcolumn}   
\usepackage{subfig}
\usepackage{float}
\usepackage{ytableau}
\newcommand\ltap{\
  \raise.3ex\hbox{$<$\kern-.75em\lower1ex\hbox{$\sim$}}\ }
\newcommand\gtap{\
  \raise.3ex\hbox{$>$\kern-.75em\lower1ex\hbox{$\sim$}}\ }

\newcommand\simge{\mathrel{%
   \rlap{\raise 0.511ex \hbox{$>$}}{\lower 0.511ex \hbox{$\sim$}}}}
\newcommand\simle{\mathrel{
   \rlap{\raise 0.511ex \hbox{$<$}}{\lower 0.511ex \hbox{$\sim$}}}}

\newcommand{\slashchar}[1]%
        {\kern .25em\raise.18ex\hbox{$/$}\kern-.75em #1}
\def\lsim{\mathrel{\raise.3ex\hbox{$<$\kern-.75em\lower1ex\hbox{$\sim$}}}}
\def\gsim{\mathrel{\raise.3ex\hbox{$>$\kern-.75em\lower1ex\hbox{$\sim$}}}}
\newcommand{\bs}{\boldsymbol}

\newcommand\CH{{\cal H}}

\newcommand\CL{{\cal L}}

\newcommand\CO{{\cal O}}

\newcommand\be{\begin{equation}}
\newcommand\ee{\end{equation}}
\newcommand\bea{\begin{eqnarray}}
\newcommand\eea{\end{eqnarray}}
\newcommand\ba{\begin{array}}
\newcommand\ea{\end{array}}
\newcommand\nn{\nonumber}
\newcommand\tx{\textstyle}

\newcommand{\thalf}{\textstyle{\frac{1}{2}}}

\newcommand{\tthird}{\textstyle{\frac{1}{3}}}
\newcommand{\tfourth}{\textstyle{\frac{1}{4}}}

\newcommand\gev{{\rm GeV}}
\newcommand\tev{{\rm TeV}}

\newcommand\lvac{\langle \Omega \vert}
\newcommand\rvac{\vert \Omega \rangle}

\newcommand\suc{SU(3)_C}

\newcommand\sutc{SU(N_{TC})}

\newcommand\atc{\alpha_{TC}}

\newcommand\Metc{M_{ETC}}

\newcommand\tom{\omega_{T}}
\newcommand\tro{\rho_{T}}

\newcommand\tpi{\pi_T}

\begin{document}

\title{
\vskip -15mm
{\Large{\bf A Composite Higgs Model with Minimal Fine-Tuning I.~The Large-$N$
    and Weak-technicolor Limit}}\\
} \author{
  {\large Kenneth Lane\thanks{lane@physics.bu.edu}}\\
  {\large Department of Physics, Boston University}\\
  {\large 590 Commonwealth Avenue}\\
  {\large Boston, Massachusetts 02215, USA}\\
} \maketitle

\begin{abstract}
  
  We suggest a criterion to minimize the amount of fine-tuning in a composite
  Higgs model. The paradigm of this type of model is the top-condensate model
  of Bardeen, Hill and Lindner (BHL). Although ``minimally fine-tuned'', this
  model failed to account correctly for the masses of the top quark and the
  125~GeV Higgs boson. We propose a generalization of the BHL model that
  employs finely-tuned extended technicolor (ETC) plus technicolor (TC)
  interactions.  The additional freedom of this model may accommodate both
  $m_t(173)$ and $M_H(125)$. This paper studies the large-$N_{TC}$ and $N_C$
  limit of this model in which technicolor is weak and does not contribute to
  electroweak symmetry breaking. Refinements including walking-TC dynamics
  and a renormalization group analysis of $m_t$ and $M_H$ will appear in a
  subsequent paper. A likely generic signal of this model is enhanced
  production of longitudinally-polarized weak bosons, alone and in association
  with $H(125)$
 
 \end{abstract}


\newpage

\section*{1. Introduction and Plan}

The concept of naturalness in particle physics is over 40 years
old~\cite{Wilson:1970ag,'tHooft:1979bh} and the quest for a natural theory of
electroweak symmetry breaking (EWSB) has been a dominant theme for almost
that long~\cite{Weinberg:1979bn,Susskind:1978ms}. This is because the {\em
  elementary} Higgs boson~\cite{Englert:1964et,Higgs:1964pj,Guralnik:1964eu}
introduced to trigger gauge symmetry breaking and give mass to the $W$ and
$Z$ in the standard electroweak model~\cite{Weinberg:1967tq, Salam:1968rm} is
so very unnatural. There is no cut-off to the quadratically divergent
corrections to its squared mass this side of the Planck scale.
The discovery at the CERN LHC of a $125\,\gev$ Higgs boson,
$H(125)$~\cite{Aad:2012tfa, Chatrchyan:2012ufa}, possibly the lone Higgs
boson of the standard model, has left supersymmetric and composite models of
a light Higgs boson as the only remaining approaches to naturalness. Both
involve a new energy scale $\Lambda$ --- either the scale of supersymmetry
breaking or the scale of the new strong dynamics binding the composite Higgs
--- that serves to cut off the corrections to $M_H^2$ at
$\CO(\Lambda^2/16\pi^2)$ or, perhaps, $\CO(\Lambda^2/(16\pi^2)^2)$. Thus,
$\Lambda$ (or $\Lambda/4\pi$) must not be larger than about 1~TeV in order
that the theory is natural.  Generally, this is achieved by having the
standard quadratic divergence in $M_H^2$ from the top quark (and weak bosons)
canceled by contributions from partners of the top (and $W,Z$). The failure,
so far, to find these partners at masses below 1-2~TeV\footnote{{\tt
    https://twiki.cern.ch/twiki,bin/view/CMSPublic/PhysicsResultsB2G}, {\tt
    https:/twiki.cern.ch/twiki/bin/view/CMSPublic/ResultsSUS}, \hfil\break
  {\tt https:/twiki.cern.ch/twiki/bin/view/ATLASPublic/ExoticsPublicResults},
  {\tt
    https:/twiki.cern.ch/twiki/bin/view/ATLASPublic/SupersymmetricPublicResults}.
}, has put considerable stress on both supersymmetric and composite Higgs
models. All such models and, in particular, composite Higgs ones --- the
subject of this paper --- require a degree of fine-tuning of parameters that
calls their ``naturalness'' into serious question~\cite{Giudice:2013yca,
  Bellazzini:2014yua}

Therefore, in order to maintain the hypothesis that $H(125)$ is a
fermion-antifermion composite, I will provisionally adopt a ``principle of
least unnaturalness'': {\em the least unnatural description of a composite
  $H(125)$ is one that involves the smallest fine-tuning for $M_H^2$ {\em
    and} the fewest number of free parameters that must be fine-tuned to
  achieve this.}

The paradigm of this sort of light composite Higgs description is the
topcolor model of Bardeen, Hill and Lindner (BHL)~\cite{Bardeen:1989ds}. In
their model, $q_L = (t,b)_L$ and $t_R$ are assumed to have a new strong
(presumably broken gauge) interaction at some high scale $\Lambda$, giving
rise to the $SU(2)\otimes U(1)$-invariant four-fermion interaction $\CL_{\bar
  tt}$ at energies below $\Lambda$,
\be\label{eq:LBHL}
\CL_{\bar tt} = G\, \bar q^{ia}_L\, t_{Ra} \,\, \bar t^b_R \,q_{Lib} \,.
\ee
Here, the $SU(2)_{EW}$ and color-$SU(3)_C$ indices, $i$~and $a,b$, are summed
over; the coupling $G = \CO(1/\Lambda^2)$. This Nambu--Jona-Lasinio (NJL)
interaction produces the top-quark mass $m_t$ {\em and} a $\bar q_L t_R$
composite scalar doublet $\phi$ if $G$ satisfies
\be\label{eq:NJL}
\frac{G N_C}{8\pi^2}\left(\Lambda^2 -
  m_t^2\ln \frac{\Lambda^2}{m_t^2}\right) = 1,\,\,\,{\rm i.e.,} 
\,\,\, G > G_c = \frac{8\pi^2}{N_C \Lambda^2}\,.
\ee
Here, $\Lambda$ is also the cutoff of the momentum integral defining $m_t$ in
the NJL-bubble, or large-$N_C$, approximation to its gap equation, while $N_C
= 3$ is the number of ordinary colors.\footnote{In the UV-complete model with
  Lagrangian $\CL_{\bar tt}$ there is no quadratic divergence from the $HHWW$
  vertex with a $W$-loop. Quadratic divergences involving weak-boson exchange
  do occur in subleading order in $1/N_C$.} The composite scalar is a complex
doublet under $SU(2)_{EW}$. It consists of three massless Goldstone bosons,
eaten by $W$ and $Z$, and a Higgs boson $H$ of mass $M_H = 2m_t$.

It is clear from Eq.~(\ref{eq:NJL}) that $m_t$ and $M_H$ can be nonzero but
very much less than $\Lambda$ if and only if $G$ is greater than but very
close to $G_c$.  This is the fine-tuning of the BHL model, but it is the
model's {\em\underbar{only}} fine-tuning. Once it is imposed, all other
$\Lambda$-dependence is logarithmic. Thus, even though $\Lambda$ is very
large in BHL, the model exemplifies our notion of being least unnatural.

The low-energy Lagrangian describing $H$ interactions with $q_L$, $t_R$ and the
EW gauge bosons is just the standard-model Lagrangian~\cite{Bardeen:1989ds}.
In that formulation, the negative $\CO(\Lambda^2)$ contribution to $M_H^2$
and the $\ln \Lambda^2$ contribution to its quartic self-interaction are
induced by the Yukawa interaction $\Gamma_t\, \bar tt H$, where $\Gamma_t$ is
obtained from the residue of the Higgs pole in the $\bar tt \to \bar tt$
amplitude in the $0^+$ channel. Then, the Higgs vacuum expectation value
(vev)~$v$ is determined in the usual way from the quartic scalar coupling
$\lambda \sim \Gamma_t^4 N_C \ln\Lambda^2/16\pi^2$ and the negative $M_H^2$.
The value of $v$ is set by $M_W = \thalf gv$, and then $m_t = \Gamma_t
v/\sqrt{2}$. Thus, $m_t$, $M_H$ and $M_{W,Z}$ are all closely related in the
BHL model. It is the most minimal {\em dynamical} model of electroweak
symmetry breaking.

The renormalization group equations for the Yukawa coupling $\Gamma_t$ of
$\bar tt H$, the quartic coupling $\lambda$, and the SM gauge couplings
$g_{1,2,3}$ result in a significant reduction of $m_t$ and $M_H/m_t$, with
smaller values obtained for larger $\Lambda$. Unfortunately, even for
$\Lambda = 10^{15}\,\gev$, BHL obtained $m_t = 229\,\gev$ and $M_H =
256\,\gev$. Still, the importance of the BHL model is that it suggests a
connection between the relatively large value of the top-quark mass and the
lightness of the Higgs boson.

The purpose of this work is to ameliorate the fine-tuning of the BHL model
and to obtain masses for the Higgs boson and top quark that are closer to
their measured values. I demonstrate this in a simple model of technicolor
(TC) plus {\em strong} extended technicolor (ETC). Technicolor with
weakly-coupled extended technicolor, cannot account for the large value of
$m_t$. But, if ETC is strong with its four-fermion coupling
$g_{ETC}^2/M_{ETC}^2$ finely-tuned, it can produce a large $m_t$ that is much
smaller than the mass scale $M_{ETC}$ of the ETC boson giving rise to
$\CL_{\bar tt}$~\cite{Appelquist:1988as,Takeuchi:1989qa}. This is similar to
the BHL model, but now the relevant scale, $\Lambda \simeq M_{ETC}$, is
expected to be $\CO(10)$--$\CO(100)\,\tev$, much lower than the BHL
scale.\footnote{Relatively low masses for ETC bosons generating
  third-generation masses need not conflict with limits on flavor-changing
  neutral current interactions.}  Furthermore, as shown in
Ref.~\cite{Chivukula:1990bc}, the symmetry breaking phase transition must be
second order. But, then, it implies the existence of a composite
complex-scalar doublet $\phi$ with three Goldstone bosons and a massive but
light scalar that couples strongly to the top quark, {\em exactly as in the
  BHL model}. This scalar will be our candidate for $H(125)$. In our model,
the Higgs boson is a composite of $\bar tt$ and $\bar UU$, where $U$ is the
up-component of a technifermion doublet $T = (U,D)$.

In this paper, we treat this TC-ETC model in the large-$N$ ($N_C,N_{TC}$)
limit. In a realistic model of this type, we expect TC to be strong enough to
participate in EWSB. That is, its coupling $\atc$ is near an infrared fixed
point of its $\beta$-function~\cite{Lane:1991qh,Appelquist:1997fp} and, so,
it evolves slowly~\cite{Holdom:1981rm,Appelquist:1986an,Yamawaki:1986zg,
  Akiba:1986rr} and reaches the critical value $\alpha_c$ for chiral symmetry
breaking at a scale $\Lambda_{TC}$ of order several hundred GeV. This seems
necessary to account for light quark and lepton masses induced by ETC bosons
with $M_{ETC}$ of many 100's of TeV~\cite{Eichten:1980ah}, for then the
relevant technifermion condensates at $M_{ETC}$ are enhanced by a large
anomalous dimension, $\gamma_m \simeq 1$~\cite{Cohen:1988sq}.

Including a walking $\atc$ in our analysis is a complication we will defer to
a subsequent paper. The renormalization group running down from $\Lambda$ of
fermion and Higgs masses referred to above will also be deferred. A further
simplification is that light quark and lepton masses are left out; their
inclusion is not technically difficult. The phenomenology of this TC-ETC
model will be developed in a third paper (see Sec.~6 for a brief foretaste).

This model has features that might allow it to account better for the Higgs
and top masses than the BHL model does. First, the technifermion contribution
to the composite Higgs loosens the tight connection among $m_t$, $M_W$ and
$M_H$. Second, there are now two large Yukawa couplings of the composite
Higgs to fermions, $\Gamma_t$ to $\bar tt$ and $\Gamma_U$ to $\bar UU$. The
renormalization group equations for $\Gamma_U$ will involve the strong
walking gauge coupling of technicolor when that is included in the model.
Third --- and this is a point we are uncertain about --- in addition to the
light scalar $H$ induced by fine-tuning, there may be a lightest $0^{++}$
technihadron bound state. This state could mix with $H$ and drive down its
mass, and certainly otherwise complicate the model's phenomenology. This
possibility will be considered in the third paper of this series.

In the remainder of this paper, then, we discuss the composite model in the
large-$N$, weak-TC limit. We assume that all of EWSB comes from the single
composite Higgs boson of the model, i.e., its vev is $v = 246\,\gev$. Our
development follows that in BHL. In Sec.~2 a model Lagrangian is presented
and used to calculate the dynamical masses $m_t$ and $m_U$ at the scale
$\Lambda$. The $\bar qq$ and $\bar TT$ scattering amplitudes are computed in
Sec.~3 and their scalar and pseudoscalar (Goldstone) poles are revealed. We
find that $M_H$ (at scale~$\Lambda$) generically lies between $2m_t$ and
$2m_U$. The electroweak gauge boson propagators in $\CO(g_{1,2}^2)$ and
large-$N$ approximations, including their Goldstone pole contributions, are
computed in Sec.~4. A numerical study of the $2\to 2$ scattering amplitudes
in the scalar channel and the value of $M_H$ in one fitting scheme are
presented in Sec.~5.  We also calculate $\Gamma_t$ and the Higgs vev~$v$ from
the residue of the Higgs pole in the $\bar tt \to \bar tt$ amplitude.
Section~6 includes preliminary comments on the model's bound-state spectrum
that may be of use to experimentalists. In particular, the possibility that
weak boson production is enhanced by $\tro$ and $\tom$ states is discussed.
We summarize the large-$N$, weak-TC results and what remains to be done in
Sec.~7.

There has been much previous work using the NJL mechanism to describe the
Higgs boson, including Refs.~\cite{Nambu:1988mr,Miransky:1988xi,
  Miransky:1989ds} which preceded~BHL in involving a new strong interaction
of top quarks as the dynamics of EWSB. Topcolor led to the so-called
top-seesaw models of Dobrescu and Hill~\cite{Dobrescu:1997nm} and Chivukula,
et al.~\cite{Chivukula:1998wd} and, more recently, Refs.~\cite{Fukano:2012qx,
  Fukano:2013kia}.\footnote{The last two papers contain a large bibliography
  of related work.} Bar-Shalom and collaborators proposed a ``hybrid model''
with a dynamical Higgs-like scalar plus an elementary scalar to describe
$H(125)$~\cite{Bar-Shalom:2013hda, Geller:2013dla}. They used an NJL
Lagrangian with fourth generation quarks interacting via a topcolor
interaction with scale $\Lambda \sim 1\,\tev$ to generate the dynamical
scalar. Apart from the use of NJL in the bubble approximation, these models
do not resemble ours, and the use of fourth generation quarks is
reminiscent of the top-seesaw mechanism. Finally, as this paper was being
written, there appeared one by Di~Chiara, et al., who proposed a model of
$H(125)$ based on TC and ETC, using a Lagrangian which is a truncated version
of that introduced in Sec.~2~\cite{DiChiara:2014gsa}.  This model bears no
further resemblance to ours; in particular, and among other things, strong,
fine-tuned ETC is not employed in their paper to make the Higgs boson much
lighter than the TC and ETC scales.

\section*{2. The TC-ETC Model in the Large-$N$ Approximation.}

The fermions in this model transform under electroweak $(SU(2)\otimes
U(1))_{EW}$, ordinary color $\suc$ and technicolor $\sutc$, and they are
\bea\label{eq:qT}
&& q_L = \left(\begin{array}{c} t\\b\\\end{array}\right)_L
\in ({\bs 2},\tx{\frac{1}{6}},{\bs 3}, {\bs 1})\,, \qquad
 t_R \in ({\bs 1},\tx{\frac{2}{3}},{\bs 3}, {\bs 1})\,,\quad
 b_R \in ({\bs 1},-\tx{\frac{1}{3}},{\bs 3}, {\bs 1})\,,\nn \\ \\\nn
&& T_L = \left(\begin{array}{c} U\\D\\\end{array}\right)_L
\in ({\bs 2},0,{\bs 1}, {\bs R_{TC}})\,, \qquad
 U_R \in ({\bs 1},\tx{\frac{1}{2}},{\bs 1}, {\bs R_{TC}})\,,\quad
 D_R \in ({\bs 1},-\tx{\frac{1}{2}},{\bs 1}, {\bs R_{TC}})\,,\nn
\eea
where ${\bs R_{TC}}$ is a complex representation of $\sutc$. As explained
above, light quarks and leptons are not dealt with in this paper. Likewise,
additional technifermions are not included here.

The {\em hard} masses of $t$ and $U$ are generated by ETC interactions at a
scale $\Lambda \simeq \Metc =\CO(10)$--$\CO(100)\,\tev$. At energies below
$\Lambda$, the effective interaction is taken to be a sum of terms similar to
the BHL Lagrangian, $\CL_{\bar tt}$, in Eq.~(\ref{eq:LBHL}):
\be\label{eq:LqT}
\CL_{ETC} = G_1\, \bar q^{ia}_L \,t_{Ra} \,\, \bar t^b_R \, q_{Lib}
          + G_2\, \left(\bar q^{ia}_L\, t_{Ra} \,\, \bar U^\alpha_R\,
            T_{Li\alpha} + {\rm h.c.} \right)
          + G_3\, \bar T^{i\alpha}_L\, U_{R\alpha} \,\, \bar U^\beta_R\,
          T_{Li\beta}  \,,
\ee
where the $SU(2)_{EW}$ and color-$SU(3)_C$ and $\sutc$ indices, $i$ and
$a,b$, and $\alpha,\beta$ are summed over. This interaction is to be thought
of as having been Fierzed from an ETC interaction involving left times
right-handed currents. The $\suc$ and $\sutc$ indices appearing here
therefore cannot correspond to exchange of ordinary massless color and TC
gluons.
The couplings $G_{1,2,3}$ are nominally positive and of $\CO(1/\Lambda^2)$.
In this simplest form of our TC-ETC model, the $D$-technifermion is assumed
to get no, or at least negligible, hard mass from ETC.
Then in the neglect of EW interactions, this model has an $(SU(2)_L\otimes
U(1)_R)_q \otimes (SU(2)_L\otimes U(1)_R)_T$ flavor symmetry which is
explicitly broken to $SU(2) \otimes U(1)$ by the $G_2$ term in $\CL_{ETC}$.
{\em If} $\CL_{ETC}$ generates {\em both} $\bar tt$ and $\bar UU$ condensates
{\em and} $G_2 \neq 0$, this flavor symmetry is spontaneously broken to
$U(1)$ and just three Goldstone bosons appear. We shall see in Sec.~6 that
all three $G_i$ are comparable and that $G_2$ is not weak. Therefore, there
are {\em not} three relatively light pseudo-Goldstone bosons.

It is not difficult to add terms that generate $m_D \neq 0$, but not so easy
to maintain $m_D = m_U$ in this model.\footnote{I thank Sekhar Chivukula for
  the conversation that led to this paragraph.} For example, adding
\be\label{eq:mDone}
 G_3\, \bar T^{i\alpha}_L\, D_{R\alpha} \,\, \bar D^\beta_R\,T_{Li\beta}
\ee
to $\CL_{ETC}$ gives an $(SU(2)_L\otimes SU(2)_R)_T$ invariant interaction
and $m_D = m_U$ only if $G_2 = 0$. But, then, there is an unacceptable triplet
of very light pseudo-Goldstone bosons (see Sec.~6). Adding instead
\be\label{eq:mDtwo}
 G_2\, \left(\bar q^{ia}_L\, b_{Ra} \,\, \bar D^\alpha_R\,
  T_{Li\alpha} + {\rm h.c.} \right) +  G_3\, \bar T^{i\alpha}_L\, D_{R\alpha}
 \,\, \bar D^\beta_R\,T_{Li\beta}
\ee
generates $m_b \neq 0$ as well as $m_D \neq 0$. These masses will differ from
$m_t$ and $m_U$, respectively. But that does not necessarily upset the
observed closeness of the $\rho$-parameter to one. Further analysis of such a
model is beyond our scope in this paper.

Following BHL, the gap equations for $m_t$ and $m_U$ {\em renormalized at the
  scale $\Lambda$} are (see the Appendix)~\footnote{As did BHL, we assume
  that the condensates $\langle \bar t \gamma_5 t\rangle = \langle \bar U
  \gamma_5 U\rangle = 0$.}$^,$\footnote{The gap equations approximated in
  Eqs.~(7,8) are integrals over an ETC boson propagator with mass $\Lambda$
  times the mass term in the $t$ or $U$ propagator. In a walking-$\atc$
  model, the $U$-mass term has both a dynamical piece, falling off roughly as
  $1/p$ above the technicolor scale $\Lambda_{TC}$ and a hard piece $m_U(p)$
  that is constant up to $p\simeq \Lambda \gg \Lambda_{TC}$. In accord with
  our weak-TC assumption, the dynamical piece is ignored. In any case, the
  hard-mass term will dominate the integral unless $m_U \ll
  \Lambda_{TC}^2/\Lambda$.}
\bea\label{eq:gap}
m_t &=& -\thalf G_1 \langle\bar tt\rangle -
\thalf G_2 \langle\bar UU\rangle \nn\\
    &=&  \frac{G_1 N_C m_t}{8\pi^2}\left(\Lambda^2 -
      m_t^2\ln\frac{\Lambda^2}{m_t^2}\right) 
          +\frac{G_2 N_{TC} m_U}{8\pi^2}\left(\Lambda^2 -
      m_U^2\ln\frac{\Lambda^2}{m_U^2}\right) \,;\\\nn\\
m_U &=& -\thalf G_2 \langle\bar tt\rangle -
\thalf G_3 \langle\bar UU\rangle \nn\\
    &=&  \frac{G_2 N_C m_t}{8\pi^2}\left(\Lambda^2 -
      m_t^2\ln\frac{\Lambda^2}{m_t^2}\right) 
          +\frac{G_3 N_{TC} m_U}{8\pi^2}\left(\Lambda^2 -
      m_U^2\ln\frac{\Lambda^2}{m_U^2}\right) \,.
\eea
Here, $N_{TC}$ is the dimensionality of the $T$-representation ${\bs
  R_{TC}}$. So long as $G_2 \neq 0$ --- which we assume throughout this paper
--- the independence of the $N_C$ and $N_{TC}$ imply that (just multiply
Eq.~(7) by $m_U$ and Eq.~(8) by $m_t$)
\be\label{eq:Gequal}
G_2 = G_1\frac{m_U}{m_t} = G_3\frac{m_t}{m_U}\,.
\ee
Then, Eqs.~(\ref{eq:gap}--\ref{eq:Gequal}) imply the
following generalization of the `fine-tuning condition in Eq.~(\ref{eq:NJL}):
\be\label{eq:gapb}
\frac{G_1N_C}{8\pi^2}\left(\Lambda^2 -
      m_t^2\ln\frac{\Lambda^2}{m_t^2}\right) 
          +\frac{G_3 N_{TC}}{8\pi^2}\left(\Lambda^2 -
      m_U^2\ln\frac{\Lambda^2}{m_U^2}\right) = 1\,.
\ee
The mass parameters in this weak-TC model, $m_t$, $m_U$, $M_W$, $M_H$ and
$\Lambda$ are not independent. If $m_t$ and $m_U$ are nonzero, only one of
the three $G_i$ is an independent parameter.

As in BHL, Eqs.~(7,8) contain this model's only quadratic divergences and,
for nonzero $m_t, m_U \ll \Lambda$, its only fine-tuning of parameters.  Once
Eq.~(\ref{eq:gapb}) is enforced, all other $\Lambda$-dependence is
logarithmic.

\section*{3. The $2\to 2$ Amplitudes in the Scalar and Goldstone Boson Channels}

We again follow BHL to calculate the $2\to 2$ amplitudes in the scalar and
pseudoscalar channels.
For the neutral scalar channel, there are three amplitudes
to calculate: $\bar tt \to \bar tt$, $\bar UU \to \bar UU$ and $\bar tt
\leftrightarrow \bar UU$. The effective Hamiltonian for the bubble is 
%
\be\label{eq:scalar}
\CH_{0^+} = -\tfourth G_1 \bar t^a t_a\, \bar t^b t_b -\thalf G_2 \bar t^a t_a\,
\bar U^\alpha U_\alpha - \tfourth G_3 \bar U^\alpha U_\alpha \, \bar U^\beta
U_\beta \,.
\ee
The $\bar tt \to \bar tt$ amplitude is
\bea\label{eq:ttscalar}
 && \Gamma_{0^+}^{\bar tt}(p) = -\thalf G_1 
    -(-\thalf G_1)^2 i\int d^4x \,e^{ip\cdot x}
   \lvac T(\bar t^at_a(x)\, \bar t^bt_b(0)\rvac\nn\\
 && \qquad\qquad\qquad -(-\thalf G_2)^2 i\int d^4x \,e^{ip\cdot x}
   \lvac T(\bar t^a U_\alpha(x)\, \bar U^\alpha t_a(0) \rvac
   +\cdots
\eea
The integrals are cut off at $\Lambda$ and evaluated in the Appendix. The
relation $G_2^2/G_1 = G_3$ makes this sum a geometric series,
\bea\label{eq:ttsum}
\Gamma_{0^+}^{\bar tt}(p) &=& -\thalf G_1 \biggl[1 - {\frac{G_1
      N_C}{8\pi^2}} \left(\Lambda^2 - m_t^2
  \ln\frac{\Lambda^2}{m_t^2} \right) 
      -{\frac{G_3 N_{TC}}{8\pi^2}}\left(\Lambda^2 - m_U^2 
\ln\frac{\Lambda^2}{m_U^2}\right) \nn\\
&\quad&
-\frac{G_1 N_C (p^2-4m_t^2)}{16\pi^2} 
\int_0^1 dx\, \ln\left(\frac{\Lambda^2}{m_t^2 - p^2 x(1-x)}\right)\nn\\
&\quad&
-\frac{G_3 N_{TC} (p^2-4m_U^2)}{16\pi^2} 
\int_0^1 dx\, \ln\left(\frac{\Lambda^2}{m_U^2 - p^2 x(1-x)}\right)
\biggr]^{-1}\,.
\eea
By Eq.(\ref{eq:gapb}), the first line on the left is zero and
$\Gamma_{0^+}^{\bar tt}$ becomes
\bea\label{eq:ttplus}
\Gamma_{0^+}^{\bar tt}(p) &=& 
m_t^2\biggl[\frac{N_C m_t^2(p^2-4m_t^2)}{8\pi^2} 
\int_0^1 dx\, \ln\left(\frac{\Lambda^2}{m_t^2 - p^2 x(1-x)}\right)\nn\\
&\qquad&
+\frac{N_{TC} m_U^2(p^2-4m_U^2)}{8\pi^2} 
\int_0^1 dx\, \ln\left(\frac{\Lambda^2}{m_U^2 - p^2 x(1-x)}\right)
\biggr]^{-1}\,.
\eea
The scalar-channel amplitudes for $\bar UU\to \bar UU$ and $\bar tt
\leftrightarrow \bar UU$ are $\Gamma_{0^+}^{\bar UU} = (m_U/m_t)^2
\Gamma_{0^+}^{\bar tt}$ and $\Gamma_{0^+}^{\bar tt \leftrightarrow \bar UU} =
(m_U/m_t)\Gamma_{0^+}^{\bar tt}$. Then the sum of the four $2\to 2$
amplitudes in the neutral scalar channel is
\bea\label{eq:zeroplus}
\Gamma_{0^+}(p) &=& 
(m_t+m_U)^2\biggl[\frac{N_C m_t^2(p^2-4m_t^2)}{8\pi^2} 
\int_0^1 dx\, \ln\left(\frac{\Lambda^2}{m_t^2 - p^2 x(1-x)}\right)\nn\\
&\qquad&
+\frac{N_{TC} m_U^2(p^2-4m_U^2)}{8\pi^2} 
\int_0^1 dx\, \ln\left(\frac{\Lambda^2}{m_U^2 - p^2 x(1-x)}\right)
\biggr]^{-1}\,.
\eea
The scalar amplitude has a pole at $p^2 = M_H^2$, the solution of
\bea\label{eq:Hmass}
&& N_C m_t^2 (M_H^2-4m_t^2)\, 
\int_0^1 dx\, \ln\left(\frac{\Lambda^2}{m_t^2 - M_H^2 x(1-x)}\right)\nn\\
&& + N_{TC} m_U^2 (M_H^2-4m_U^2)
\int_0^1 dx\, \ln\left(\frac{\Lambda^2}{m_U^2 - M_H^2 x(1-x)}\right)  = 0\,.
\eea
This result, the Higgs mass at scale $\Lambda$ in the large-$N$
approximation, will be modified by renormalization-group running from
$\Lambda$ down to the $H$-pole. The BHL mass, $M_H(\Lambda) = 2m_t$, is
obtained by setting $m_U = 0$ in Eq.~(\ref{eq:Hmass}). A very good
approximation to the solution to Eq.~(\ref{eq:Hmass}) is
\be\label{eq:Hmassb}
M_H = 2\sqrt{\frac{N_C m_t^4 + N_{TC} m_U^4}{N_C m_t^2 + N_{TC} m_U^2}}
\ee
The effective Hamiltonian for the neutral and charged Goldstone poles in
$2\to 2$ scattering is
\bea\label{pseudo}
\CH_{0^-} &=&  \tfourth G_1 \bar t^a \gamma_5 t_a\, \bar t^b \gamma_5 t_b +
\thalf G_2 \bar t^a \gamma_5 t_a\, \bar U^\alpha \gamma_5 U_\alpha + 
\tfourth G_3 \bar U^\alpha \gamma_5 U_\alpha \, \bar U^\beta \gamma_5 U_\beta  
\nn\\
&\quad&  -\, G_1\, \bar b^{a}_L \,t_{Ra} \, \bar t^b_R \, b_{Lb}
          - G_2\, \left(\bar b^{a}_L\, t_{Ra} \, \bar U^\alpha_R\,
            D_{L\alpha} + {\rm h.c.} \right)
          - G_3\, \bar D^{\alpha}_L\, U_{R\alpha} \, \bar U^\beta_R\,
          D_{L\beta} \,.
\eea
The $\bar tt \to \bar tt$ amplitude is (note the $i$'s in $i\gamma_5$, left
out in BHL):
\be\label{eq:ttGB}
 \Gamma_{0^-}^{\bar tt}(p) = -\thalf G_1 
    -(\thalf G_1)^2 i\int d^4x \,e^{ip\cdot x}
   \lvac T(\bar t^a i\gamma_5 t_a(x)\, \bar t^b i\gamma_5 t_b(0)\rvac +\cdots\,.
\ee
Proceeding as in the scalar case. the sum of the $2\to 2$ amplitudes in the
neutral channel is
\bea\label{eq:neutral}
\Gamma^0_{0^-}(p) &=& \frac{8\pi^2 (m_t+m_U)^2}{p^2}
\biggl[N_C m_t^2 \int_0^1 dx\, \ln\left(\frac{\Lambda^2}{m_t^2 - p^2
    x(1-x)}\right)\nn\\ 
&\qquad&
+ N_{TC} m_U^2 \int_0^1 dx\, \ln\left(\frac{\Lambda^2}{m_U^2 - p^2
    x(1-x)}\right) \biggr]^{-1}\,.
\eea
The corresponding amplitude in the charged $t\bar b \to t \bar b$ channel is
\be\label{eq:tbGBa}
 \Gamma_{0^-}^{t\bar b}(p) = -\tfourth G_1 
    -(G_1)^2 i\int d^4x \,e^{ip\cdot x}
   \lvac T(\bar t_R^a b_{La}(x)\, \bar b_L^b t_{Rb}(0)\rvac +\cdots\,.
\ee
Including the other channels, these sum up to
\bea\label{eq:tbGB}
\Gamma^{\pm}_{0^-}(p) &=& \frac{2\pi^2 (m_t+m_U)^2}{p^2}
\biggl[N_C m_t^2 \int_0^1 dx\, x\ln\left(\frac{\Lambda^2}{m_t^2 x - p^2
    x(1-x)}\right)\nn\\ 
&\qquad&
+ N_{TC} m_U^2 \int_0^1 dx\, x\ln\left(\frac{\Lambda^2}{m_U^2 x - p^2
    x(1-x)}\right) \biggr]^{-1}\,.
\eea
The manipulations~\cite{Bardeen:1989ds} used to obtain these results are
given in the Appendix.

As noted above, there may be an isotriplet of pseudo-Goldstone bosons that
acquire mass from the $G_2$-interaction. This is discussed briefly in Sec.~6
and considered in more detail in a later paper.

\section*{4. The Electroweak Gauge Boson Propagators}

In this section we compute the EW propagators in the NJL-bubble
approximation, neglecting EW gauge-boson radiative corrections. As in BHL, the
EW fields are rescaled to bring the gauge coupling into their kinetic
terms, i.e., $(1/4g^2)F_{\mu\nu}^2$. The $(SU(2)\otimes U(1))_{EW}$ currents
are
\bea\label{eq:currents}
j_{\mu}^A &=& \bar q_L\gamma_\mu \frac{\tau_A}{2} q_L +
               \bar T_L\gamma_\mu \frac{\tau_A}{2} T_L \qquad
                    (A=1,2,3)\,;\nn\\ 
j_\mu^0 &=& {\tx{\frac{1}{6}}}\left(\bar q_L \gamma_\mu q_L +
\bar q_R \gamma_\mu q_R\right) + \bar q_R\gamma_\mu \frac{\tau_3}{2} q_R 
                    + \bar T_R\gamma_\mu \frac{\tau_3}{2} T_R\,.
\eea

The inverse $W$-propagator is
\bea\label{eq:Wprop}
\frac{1}{g_2^2}(D^\pm(p))^{-1}_{\mu\nu} &=& \frac{1}{g_2^2}(p_\mu p_\nu - p^2
g_{\mu\nu}) + 
\frac{i}{2}\int d^4x\,e^{ip\cdot x} \lvac T(j_{L\mu}^{(1+i2)}(x)\,
j_{L\nu}^{(1-i2)}(0))\rvac \nn\\
&=& (p_\mu p_\nu - p^2 g_{\mu\nu})\left(\frac{1}{g_{2W}^{2}(p^2)} -
  \frac{f_W^2(p^2)}{p^2}\right)\,.
\eea
In the second line of Eq.~(\ref{eq:Wprop}), $g_{2W}^{-2}(p^2)$ is computed
from the bare inverse $W$-propagator plus the one-loop correlator
$\Pi^\pm_{\mu\nu}(p)$ of a pair of charged weak currents. It is given in the
TC-ETC model by (see the Appendix for details)
\bea\label{eq:gtwoW}
g_{2W}^{-2}(p^2) &=& g_2^{-2} + \frac{1}{16\pi^2}\int_0^1 dx \, 2x(1-x)
    \biggl[N_C \ln\biggl(\frac{\Lambda^2}{m_t^2 x - p^2 x(1-x)}\biggr) \nn\\
&+&  N_{TC} \ln\biggl(\frac{\Lambda^2}{m_U^2 x - p^2 x(1-x)}\biggr)\biggr]\,.
\eea
The contribution to the $W$-propagator from the massless pole in
$\Gamma^{\pm}_{0^-}$ in Eq.~(\ref{eq:tbGB}) is
\bea\label{eq:fW}
f_W^2(p^2) &=& \frac{1}{16\pi^2}\int_0^1 dx\, x
    \biggl[N_C\,m_t^2 \ln\biggl(\frac{\Lambda^2}{m_t^2 x - p^2 x(1-x)}\biggr)
    \nn\\
&+&  N_{TC}\,m_U^2 \ln\biggl(\frac{\Lambda^2}{m_U^2 x - p^2
  x(1-x)}\biggr)\biggr]\,.
\eea

A comment is in order here: The fermion masses in the one-loop-EW and
fermion-bubble-sum contributions to the weak-boson propagators must be the
hard masses generated by $\CL_{ETC}$. It is these masses that satisfy the gap
Eqs.~(7,8), and those relations are used to remove the $\Lambda^2$-dependence
from the bubble sums. Furthermore, the masses in the $m^2 g_{\mu\nu}$ part of
the EW loop must be the same as those in the $m^2 p_\mu p_\nu/p^2$ terms
coming from the bubble sum in order that Ward-Takahashi (WT) identities are
maintained and the propagators are transverse. Therefore, in accord with the
model defined by $\CL_{ETC}$, $m_b = M_D = 0$ in $g_{2W}^2(p^2)$ and
$f_W^2(p^2)$. A more complete treatment of the propagators will include the
dynamical strong-TC contributions to the Goldstone poles and, to satisfy the
WT identities, the fermion masses. These do not have $\ln\Lambda^2$
dependence as they are cut-off by TC dynamics (mainly) at scale
$\Lambda_{TC}$. This more complicated analysis is deferred to a later paper.
The upshot of all this is that in the weak-TC limit all of EWSB comes from
$\CL_{ETC}$ and, since there is just one complex Higgs doublet,
\be\label{eq:zip}
f_W^2(0) = 1/(4\sqrt{2}G_F) = (123\,\gev)^2\ \cong M_W^2/g_{2W}^2(0),.
\ee
%
%
  




The inverse neutral propagator matrix is
\bea\label{eq:Zpropa}
\frac{1}{g_ig_j}(D^0(p))^{-1}_{\mu\nu} &=& \left(\begin{array}{cc}
1/g_2^2 & 0\\ 0 & 1/g_1^2\end{array}\right)
(p_\mu p_\nu - p^2 g_{\mu\nu})\nn\\
&+& i\int d^4x\,e^{ip\cdot x} \lvac \left(\begin{array}{cc}
T(j_\mu^3(x) j_\nu^3(0)) & T(j_\mu^3(x) j_\nu^0(0))\\
T(j_\mu^0(x) j_\nu^3(0)) & T(j_\mu^0(x) j_\nu^0(0))\end{array}\right)\rvac\,.
\eea
As for the $W$-propagator, this is calculated from the bare inverse
propagators, the one-loop neutral-current correlators, and the large-$N$ bubble
sums for $\Gamma^0_{0^-}$. This gives
\be\label{eq:Zprop}
\frac{1}{g_ig_j}(D^0(p))^{-1}_{\mu\nu} = (p_\mu p_\nu - p^2 g_{\mu\nu})
\left(\begin{array}{cc}1/g_{2Z}^2(p^2) & 0\\ 0
    &1/g_{1Z}^2(p^2)\end{array}\right)
- \left(\begin{array}{cc} 1 & -1\\ -1& 1\\
  \end{array}\right)\frac{f_Z^2(p^2)}{p^2} \,,
\ee
where $g_{2Z}^2(p^2)$, $g_{1Z}^2(p^2)$, and $f_Z^2(p^2)$b have been defined
to give a massless photon pole in the diagonalized neutral propagator.
Reading off $f_Z^2$ from the massless pole term in the $\langle j_\mu^3
j_\nu^0\rangle$-correlator, we obtain (see the Appendix)
\bea\label{eq:gtwoZ}
f_Z^2(p^2) &=& \frac{1}{32\pi^2}\int_0^1 dx\,
    \biggl[N_C\,m_t^2 \ln\biggl(\frac{\Lambda^2}{m_t^2 - p^2 x(1-x)}\biggr)
 + N_{TC}\,m_U^2 \ln\biggl(\frac{\Lambda^2}{m_U^2 - p^2
   x(1-x)}\biggr)\biggr]\nn\\
&+& \frac{N_C p^2}{16\pi^2}\int_0^1dx\, {\tthird} x(1-x)
\ln\biggl(\frac{-p^2 x(1-x)}{m_t^2 - p^2 x(1-x)}\biggr)
\,;\\\nn\\
g_{2Z}^{-2}(p^2) &=& g_2^{-2} 
+ \frac{1}{16\pi^2}\int_0^1 dx \, x(1-x)
\biggl\{N_C\biggl[{\tx{\frac{4}{3}}}\ln\biggl(\frac{\Lambda^2}{m_t^2 -
    p^2 x(1-x)}\biggr)
+ {\tx{\frac{2}{3}}}\ln\biggl(\frac{\Lambda^2}{-p^2 x(1-x)}\biggr)\biggr]\nn\\ 
&+& N_{TC}\biggl[\ln\biggl(\frac{\Lambda^2}{m_U^2  - p^2 x(1-x)}\biggr)
+ \ln\biggl(\frac{\Lambda^2}{-p^2 x(1-x)}\biggr)\biggr]\biggr\}\,; \\\nn\\
g_{1Z}^{-2}(p^2) &=& g_1^{-2} 
+ \frac{1}{16\pi^2}\int_0^1 dx \, x(1-x)
\biggl\{N_C\biggl[{\tx{\frac{20}{9}}}\ln\biggl(\frac{\Lambda^2}{m_t^2  -
    p^2 x(1-x)}\biggr)
+ {\tx{\frac{2}{9}}}\ln\biggl(\frac{\Lambda^2}{-p^2
  x(1-x)}\biggr)\biggr]\nn\\
&+& N_{TC}\biggl[\ln\biggl(\frac{\Lambda^2}{m_U^2 - p^2 x(1-x)}\biggr)
+ \ln\biggl(\frac{\Lambda^2}{-p^2 x(1-x)}\biggr)\biggr]\biggr\}\,.
\eea
The $(-p^2 x(1-x))$ arguments of the logarithms come from a $b$ or
$D$-fermion loop with $m_b = m_D = 0$.

The $\rho$-parameter,
\be\label{eq:rho}
\rho \cong \frac{f_W^2(0)}{f_Z^2(0)} =
\frac{(32\pi^2)^{-1}\left[N_C\,m_t^2\left(\ln(\Lambda^2/m_t^2)+\thalf\right) +
N_{TC} \,m_U^2\left(\ln(\Lambda^2/m_U^2)+\thalf\right)\right]}
{(32\pi^2)^{-1}\left[N_C \,m_t^2 \ln(\Lambda^2/m_t^2) +
N_{TC} \,m_U^2 \ln(\Lambda^2/m_U^2)\right]} \,,
\ee
the running of the EW gauge couplings, and the $W$ and $Z$ pole masses,
solutions of
\bea\label{eq:MWMZ}
M_W^2 &=& g^2_{2W}(M_W^2)f^2_W(M_W)\,, \nn\\
M_Z^2 &=& (g^2_{1Z}(M_Z^2)+g^2_{2Z}(M_Z^2))f^2_Z(M_Z)\,,
\eea
will be discussed in the numerical calculations next, in Sec.~5.

There has been much discussion over the years of the constraint on
technicolor theories from the $S$-parameter~\cite{Peskin:1990zt,
  Golden:1990ig, Holdom:1990tc,Altarelli:1991fk}. However, as emphasized in
Refs.~\cite{Lane:1994pg,Lane:2002wv}, all of these calculations of $S$ assume
that TC dynamics is QCD-like, with asymptotic freedom setting in rather
quickly above $\Lambda_{TC}$. But TC dynamics cannot be QCD-like. As noted
earlier, $\atc$ must be a walking gauge coupling to avoid unwanted large
flavor-changing neutral current interactions, and this invalidates the
assumptions made to calculate $S$. Calculating $S$ in the strong dynamics of
walking technicolor is now the object of a number of groups using lattice
gauge theoretic techniques; see, e.g., Refs.~\cite{Appelquist:2010xv,
  Schaich:2011qz,Appelquist:2013sia}.

\section*{5. Numerical Calculations in the Large-$N$ Approximation}

 \begin{table}[!ht]
     \begin{center}{
  \begin{tabular}{|c|c|c|c|c|c|}
  \hline
 $\Lambda$ & $m_t$ & $m_U$ & $M_H$ &$\Gamma_t$ & $v=\sqrt{2} m_t/\Gamma_t$ \\
  \hline\hline
 $20\,\tev$ & $134\,\gev$ & $167\,\gev$ & $330\,\gev$ & 0.783  & $242\,\gev$ \\
 $500\,\tev$ & $118\,\gev$ & $126\,\gev$ & $250\,\gev$ & 0.685  & $244\,\gev$ \\
  \hline\hline
 $\Lambda$ & $\rho$ & $g_1$ & $g_2$ & $M_W$({\rm pole}) &$M_Z({\rm pole})$ \\
   \hline\hline
 $20\,\tev$  & 1.0520 &0.3941  & 0.7187 & 80.8  & 91.0  \\
 $500\,\tev$ & 1.0301 &0.4230  & 0.7714 & 80.6  & 93.0 \\
\hline
 \end{tabular}}
 \caption{The Higgs mass, $\rho$-parameter and the $W$, $Z$-pole masses
 calculated for ETC scales $\Lambda = 20$ and $500\,\tev$. The calculation
 scheme adopted is described in the text.\label{tab:numerical}}
 \end{center}
 \end{table}

Here we present outcomes of the large-$N$, weak-TC results of Secs.~3 and~4
for a simple numerical scheme. In this scheme we fix $\Lambda$ and then
calculate $m_t(\Lambda)$ using one-loop QCD running with six flavors from
$\Lambda$ down to $2m_t$ and with five flavors down to $m_t =
173\,\gev$~\cite{Beringer:1900zz}. Fixing $f_W(0) = (4\sqrt{2}G_F)^{-1/2}
\cong 123\,\gev$ in Eq.~(\ref{eq:rho}) then determines $N_{TC} m_U^2$. Note
that, in leading-log approximation, this same combination, $N_{TC} m_U^2$,
appears in the formula for $M_H^2$, Eq.(\ref{eq:Hmass}). Finally, we choose
$N_{TC} = 15$ (which corresponds to an $SU(6)$ TC gauge group with $T=(U,D)$
in the antisymmetric second-rank tensor representation) and calculate
$M_H(\Lambda)$. As a check on our calculation, we obtain the Higgs vev~$v$
from $m_t(\Lambda) = \Gamma_t(\Lambda) v/\sqrt{2}$, where
\be\label{eq:Gammat}
\left(\Gamma_t(\Lambda)/\sqrt{2}\right)^2 = \lim_{p^2\to M_H^2} (p^2 -
M_H^2) \Gamma_{0^+}^{\bar tt}(p)\,,
\ee
and compare the result to the input $2f_W(0) = 246\,\gev$.

We consider two cases, $\Lambda = 20\,\tev$ and $\Lambda = 500\,\tev$. There
is no obvious reason not to have such a high scale for generating $m_t$ since
the only price is more fine tuning of $G_1$ in Eq.~(\ref{eq:LqT}). In a more
complete TC-ETC model, such a large ETC mass for the third generation may
also suffice to produce masses for the lighter quarks while suppressing their
flavor-changing neutral current interactions. This would eliminate the need
for a ``tumbled'' spectrum of ETC masses. The difference in the two cases we
consider is greater than one might have anticipated given the merely
logarithmic dependence on $\Lambda$ of the $2\to 2$ amplitudes calculated in
Sec.~3.

The results are in Table~\ref{tab:numerical}. The EW couplings $g_{1,2}$ used
in Eqs.~(\ref{eq:gtwoW},31,32) to calculate the $W$ and $Z$
pole-masses were determined by requiring that $g^{-2}_{1,2Z}$ at $p = M_Z =
91.18\,\gev$ give $\sin^2\theta_W(M_Z) = g_1^2/(g_1^2 + g_2^2) =
0.23116$ and $\alpha(M_Z) = 1/128$~\cite{Beringer:1900zz}. All the results in
the lower half of the table are very good except for a slightly high $Z$-pole
mass in the second case.

It is worth recalling that the Higgs mass will be renormalized from $\Lambda$
down to the electroweak scale. In BHL~\cite{Bardeen:1989ds}, $M_H(\Lambda =
10^{15}\,\gev) = 330\,\gev$ decreased to $256\,\gev$. We expect our values of
$M_H$ to decrease as well but, of course, we cannot guess by how much ---
especially since the effect of a walking TC coupling has to be included in
the running of $m_U$. Another thing is that, since we are assuming there is
only one Higgs boson, its coupling to the top quark {\em at the top mass}
will be $v/(\sqrt{2}m_t) \cong 1$, so that the $ggH$ coupling of the Higgs to
QCD gluons will have its standard-model strength and the production rates
$\sigma(gg \to \gamma \gamma,\, ZZ^*,\, WW^*)$ will be as in the standard
model.

\section*{6. Preliminary Remarks on the Model's Phenomenology}

The low-lying bound-state spectrum of this model depends on the magnitude of
the ETC couplings $G_i$ and the TC gauge coupling
$\atc$.\footnote{Presumably, once its effects are included, $\atc$ becomes
  strong enough to confine technicolor at a distance scale of
  $\CO(1\,\tev^{-1})$.} The chiral-flavor symmetry of our model is
$(SU(2)_L\otimes U(1)_R)_q \otimes (SU(2)_L\otimes U(1)_R)_T$ explicitly
broken to $SU(2)\otimes U(1)$ by $G_2$.

Suppose first, then, that $G_2 = 0$. There are three possibilities: (1) $G_1$
and $G_3$ are supercritical, i.e., $G_1 > 8\pi^2/N_C\Lambda^2$, $G_3 >
8\pi^2/N_{TC}\Lambda^2$ and $m_t$, $m_U \neq 0$ are solutions of the gap
Eqs.~(7,8); (2) $G_3$ is supercritical ($m_U \neq 0$), but $G_1$ is not ($m_t
= 0$); and (3)~vice-versa. Possibilities (2) and (3) are excluded because (2)
$m_t$ cannot be zero and (3) TC would likely generate a dynamical mass for
$U$ and $D$, spontaneously breaking their chiral symmetry, now $SU(2)_L
\otimes SU(2)_R$, giving rise to an additional triplet of massless Goldstone
bosons. They could acquire mass only from the EW gauge interactions and they
would be very light~\cite{Eichten:1980ah}, hence excluded (e.g., by
production of the charged pair in $e^+e^-$ annihilation). In possibility~(1),
we would have two light composite scalar doublet bound states, hence two
light Higgs bosons and two massless Goldstone triplets. One combination of
the Goldstone triplets would be eaten by the~$W$ and~$Z$, but the orthogonal
triplet is again very light and excluded.

Once $G_2 \neq 0$, its magnitude is fixed by Eqs.~(\ref{eq:Gequal},
\ref{eq:gapb}). Both $m_t$ and $m_U$ are nonzero. Now, both terms in
Eq.~(\ref{eq:gapb}) must be less than one. We saw in Sec.~5 that a very small
$m_U/m_t$ or $m_t/m_U$ is unlikely to be compatible with $f_W(0) \cong
123\,\gev$.\footnote{If we relax this condition on $f_W(0)$, then there must
  be at least two Higgs doublets and, therefore, two Higgs bosons in a TC-ETC
  model. This is an interesting complication that we do not pursue in this
  paper.} Thus, $G_2$ is not much different from $G_1$ and/or $G_3$ and it
cooperates with them to make the model just barely critical, producing
$0 < m_t,\,m_U \ll \Lambda$ --- our model's fine-tuning.

What does this mean for the spectrum of relatively low-lying bound states? So
long as TC is present and confining, we expect isovector $\tro$ and isoscalar
$\tom$ which are $\bar TT$ states. It is not clear how heavy the lightest
$\tro$ and $\tom$ are. If their binding is due mainly to TC, we would guess
their masses are in the range $500\,\gev$ to $2\,\tev$. If the strong ETC
interactions $\CL_{ETC}$ contribute to their mass {\em other than through the
  hard mass $m_U$,} they might be much heavier. Because the hard
technifermion mass is an $I=1$ operator, the neutral and charged $\tro$ and
the $\tom$ should all have nearly the same mass. It is also possible that
the mass-eigenstate vectors are ideally mixed $\bar UU$ and $\bar DD$ states.



Assuming they are lighter than $\sim 2\,\tev$, the $\tro$ and $\tom$ will be
produced at the LHC at observable rates by the Drell-Yan process, $\bar qq
\to \gamma,Z,W \to \tro$ or $\tom$~\cite{Lane:2002sm} and, if they are heavy
enough, via weak vector boson fusion (see Delgado, Grojean, Maina and
Rosenfeld in Ref.~\cite{Brooijmans:2008se}).

How do $\tro$ and $\tom$ decay? There may be a triplet of lightest
``pseudoscalars'', induced by the criticality of $\CL_{ETC}$ and by TC. This
triplet would be an admixture of $\bar qt$ and $\bar TU$ states that is
orthogonal to the three Goldstone bosons eaten by $W^\pm$ and $Z^0$.  They
are not light pseudo-Goldstone bosons, for they get a large mass from the
near-critical $G_2$ interaction. In fact, there is no obvious reason that
they are much lighter than $\tro$ and $\tom$. Thus, we expect the vectors'
dominant decay modes to involve {\em longitudinally-polarized} weak bosons,
the erstwhile ``pions'' absorbed in the Higgs mechanism, alone and possibly
in association with $H(125)$:
\bea\label{eq:Tdecays}
\rho_T^{\pm,0} &\to& W^\pm_L Z_L,\,\, W^+_L W^-_L \quad {\rm and} \quad
W^\pm_L H,\,\, Z_L H\,; \\
\tom &\to& W^+_L W^-_L Z_L\ \quad {\rm and} \quad Z_L H\,.
\eea
These decays are strong (TC) interactions. Thus, heavier $\tro$ and $\tom$
are unlikely to be narrow resonances. In that case, the presence of the
$\tro$ and $\tom$ will be signaled by increases in the rates of the above
processes at higher invariant masses.


\section*{7. Summary and Plans}

In this paper we presented a simple model of a light composite Higgs boson.
It is inspired by the top-condensate model of Bardeen, Hill and
Lindner~\cite{Bardeen:1989ds} and, in our view, its paradigmatic position as
a dynamical model embodying our notion of ``least unnaturalness''. Our model
combines technicolor with strong extended technicolor to jointly account for
electroweak symmetry breaking, the light Higgs boson discovered at the LHC,
and the mass of the top quark. The strong ETC interaction with finely-tuned
couplings is essential for these to occur at energies much less than the ETC
scale~$\Lambda$. This mechanism was anticipated in
Ref.~\cite{Chivukula:1990bc}.

Our simple model employs one technifermion doublet $T=(U,D)$ interacting with
the third generation quarks $q = (t,b)$ via three ETC interactions with
strengths $G_1,G_2,G_3 = \CO(1/\Lambda^2)$, where $\Lambda \sim
10$--$500\,\tev$. These interactions were treated in the NJL approximation of
large $(N_C,N_{TC})$.  While the TC interaction of $T$ is expected to be an
important part of the model, it is also a significant complication. We
neglected TC in this paper.

The solution of the model in this large-$N$, weak-TC limit then closely
followed BHL: The gap equations in Sec.~2 for the hard masses $m_t$ and $m_U$
are quadratically divergent, and requiring $m_t, m_U \ll \Lambda$ is a fine
tuning of a part in $\CO(\Lambda^2/m^2)$. These gap equations also imply the
relation $G_2 = G_1(m_U/m_t) = G_3(m_t/m_U)$ among the model's ETC couplings.
This relation was essential for turning the complicated NJL bubble sums for
the $2\to 2$ scattering amplitudes in Sec.~3 into simple geometric series. As
in BHL, all $\Lambda^2$-dependence in these amplitudes was removed by
applying the condition~(\ref{eq:gapb}) for nontrivial solutions to the gap
equations. In the scalar channel, the Higgs boson pole occurs at $M_H^2 \cong
4(N_C m_t^4 + N_{TC} m_U^4)/(N_C m_t^2 + N_{TC} m_U^2)$. This is the model's
only Higgs boson (necessarily, since $G_2$ is not weak), so its vev is $v =
246\,\gev$.  There are three Goldstone boson channels. Their massless poles
disappear from the physical spectrum, producing the the massive~$W$
and~$Z$-boson poles in their propagators (Sec.~4). Integrals used in Secs.~3
and~4 are in the Appendix.

In Sec.~5 we carried out a simple numerical analysis of our model by
(1)~fixing $\Lambda$ and then $m_t(\Lambda)$ so that $m_t = 173\,\gev$ at the
weak scale and (2)~determining $N_{TC}m_U^2(\Lambda)$ from the residue of the
Goldstone pole in the $W$-propagator, $f_W(0) = 123\,\gev$. Fixing
$N_{TC}$ then determined $M_H(\Lambda)$. The results are in
Table~\ref{tab:numerical} for the two choices $\Lambda = 20\,\tev$ and
$500\,\tev$. As in BHL, these values of the Higgs mass are expected to
decrease when run down to the weak scale. However, the effect of TC on the
running is unknown and, like the inclusion of TC dynamics, is deferred to the
next paper. The $\rho$-parameter and the $W$ and $Z$-pole masses were also
calculated and in quite good agreement with experiment. Of course, more
elaborate numerical schemes are possible, e.g., a ``best fit'' to $m_t$ and
$M_H$ with fixed values of $\Lambda$ and the Higgs vev $v = 2 f_W(0) =
246\,\gev$, or even a scan over $\Lambda$ for a best fit to $m_t$ and $M_H$.

Finally, in Sec.~6 we speculated briefly on the model's phenomenology. We
noted that the model with $G_2 = 0$ is excluded because it has a triplet of
nearly massless pseudo-Goldstone bosons; the charged ones would have been
discovered decades ago in $e^+e^-$ annihilation. Further, the constraint
$f_W(0) = 123\,\gev$ implies that $m_t$ and $m_U$ are likely to be comparable
and this, in turn, implies that all three $G_i$ are comparable and {\em
  nearly} critical, i.e., nearly large enough to induce nonzero $m_t,m_U$ by
themselves. Given this, it is difficult to see what this model's
phenomenology is because it will be controlled by two strong interactions, TC
and ETC, with very different energy scales. One possibility that suggested
itself deals with the model's lowest lying spin-one, isovector and isoscalar
$\bar TT$ states. If their binding is determined by TC, not ETC, dynamics,
the masses of these $\tro$ and $\tom$ should be $\thalf$--$2\,\tev$ and
possibly within reach of the LHC. Their spin-zero $\tpi$ partners are not
pseudo-Goldstone bosons and, so, are likely to be as heavy as they are.
Then, the principal observational modes of the vectors are their {\em strong}
decays to longitudinally polarized weak bosons, either in diboson and
triboson combinations or in association with $H(125)$. Beyond this,
understanding the phenomenology of this model, or any model like it, requires
a much better understanding of its dynamics. This and the phenomenology are
the subjects of planned papers.

\vfil\eject

\section*{Appendix: Integrals used in the text}

The calculations presented here come from Ref.~\cite{Bardeen:1989ds} and
C.~T.~Hill (private communication). Momentum integrals are in Minkowski space
until Wick-rotated and then cut off at momentum~$\Lambda$.

\subsection*{Sections 2 and 3:}

\noindent The fermion condensates at scale $\Lambda$ in Eq.~(\ref{eq:gap}):
\bea\label{ttcond}
\langle\bar tt\rangle_\Lambda &=& \sum_a\langle\bar
t^a(0)t_a(0)\rangle_\Lambda = -iN_C ({\rm Tr} S_t(0))_\Lambda \nn\\
&\equiv& -4iN_C\int^\Lambda \frac{d^4k}{(2\pi)^4} \frac{m_t}{k^2-m_t^2} = 
-\frac{N_C m_t}{4\pi^2}\int_0^{\Lambda^2} d k^2 \frac{k^2}{k^2 + m_t^2} \nn\\
&\cong& -\frac{N_C m_t}{4\pi^2}\left[\Lambda^2 - m_t^2\ln(\Lambda^2/m_t^2)\right]\,,
\eea
for $\Lambda^2 \gg m_t^2$.

\noindent The scalar $t\bar t \to t\bar t$ integral in Eq.~(\ref{eq:ttscalar}):

\bea\label{scalartt}
&& i\int d^4x\, e^{ip\cdot x}\lvac T(\bar t^a t_a(x)\, \bar t^b t_b(0))\rvac \nn\\
&& \quad = 4iN_C \int^\Lambda \frac{d^4k}{(2\pi)^4} \frac{k\cdot(k+p)+m_t^2}
{((k+p)^2-m_t^2)(k^2-m_t^2)} \nn\\
&& \quad = 2iN_C \int^\Lambda \frac{d^4k}{(2\pi)^4}\frac{((k+p)^2-m_t^2) +
  (k^2-m_t^2) - (p^2-4m_t^2)}{((k+p)^2-m_t^2)(k^2-m_t^2)}\nn\\
&& \quad \cong \frac{N_C}{4\pi^2}\left[\Lambda^2 -
  m_t^2\ln(\Lambda^2/m_t^2)\right] + 
\frac{N_C(p^2-4m_t^2)}{8\pi^2} \int_0^1 dx \ln\left(\frac{\Lambda^2}{m_t^2 -
p^2x(1-x)}\right)\,.
\eea
 
\noindent The Goldstone boson $t \bar t \to t \bar t$ integral in
Eq.~(\ref{eq:ttGB}):

\bea\label{eq:GBtt}
&& i\int d^4x\, e^{ip\cdot x}\lvac T(\bar t^a i\gamma_5 t_a(x)\, \bar t^b
i\gamma_5 t_b(0))\rvac \nn\\
&& \quad= 4iN_C \int^\Lambda \frac{d^4k}{(2\pi)^4} \frac{k\cdot(k+p)-m_t^2}
{((k+p)^2-m_t^2)(k^2-m_t^2)} \nn\\
&& \quad = 2iN_C \int^\Lambda \frac{d^4k}{(2\pi)^4}\frac{((k+p)^2-m_t^2) +
  (k^2-m_t^2) - p^2}{((k+p)^2-m_t^2)(k^2-m_t^2)}\nn\\
&& \quad \cong \frac{N_C}{4\pi^2}\left[\Lambda^2 -
  m_t^2\ln(\Lambda^2/m_t^2)\right] + 
\frac{N_C p^2}{8\pi^2} \int_0^1 dx \ln\left(\frac{\Lambda^2}{m_t^2 -
p^2x(1-x)}\right)\,.
\eea

\vfil\eject

\noindent The Goldstone boson $t \bar b \to t \bar b$ integral in
Eq.~(\ref{eq:tbGB}):

\bea\label{GBtb}
&& i\int d^4x\, e^{ip\cdot x}\lvac T(\bar t_R^a b_{La}(x)\, \bar b_L^b
t_{Rb}(0))\rvac \nn\\
&& \quad= 2iN_C \int^\Lambda \frac{d^4k}{(2\pi)^4} \frac{k\cdot(k+p)}
{(k+p)^2 (k^2-m_t^2)} 
 = 2iN_C \int^\Lambda \frac{d^4k}{(2\pi)^4}\frac{(k+p)^2 - k\cdot p -
  p^2} {(k+p)^2 (k^2-m_t^2)}\nn\\
&& \quad \cong \frac{N_C}{8\pi^2}\left[\Lambda^2 -
  m_t^2\ln(\Lambda^2/m_t^2)\right] 
 + \frac{N_C p^2}{8\pi^2}\int_0^1 dx\, x \int_0^{\Lambda^2}
\frac{k^2}{k^2 + m_t^2 x - p^2 x(1-x)} \nn\\
&& \quad \cong \frac{N_C}{8\pi^2}\left[\Lambda^2 -
  m_t^2\ln(\Lambda^2/m_t^2)\right] +
\frac{N_C p^2}{8\pi^2} \int_0^1 dx\, x \ln\left(\frac{\Lambda^2}{m_t^2 x -
p^2x(1-x)}\right)\,.
\eea

\subsection*{Section 4:}

There are two types of terms in this calculation, the correlator of two weak
currents and the sum of large-$N$ bubbles inserted into this correlator. The
first is the standard one-loop correction to the weak polarization tensors
$\Pi_{\mu\nu}(p)$, while the second produces the Goldstone pole in this
one-loop term. We calculate the terms here for the product of two charged
$tb$ currents. For the simple one-loop correlator of two {\em conserved}
currents, we use dimensional regularization to avoid a spurious quadratic
divergence. With $d = 4-\epsilon$ and $m_b = 0$, we have for the $t\bar b \to
t\bar b$ contribution:
\bea\label{eq:Pitb}
\Pi^{\pm}_{\mu\nu}(p) &=& \frac{i}{g_2^2}\left(\frac{g_2}{\sqrt{2}}\right)^2
\int d^4x\, e^{ip\cdot x}\lvac T(\bar t_L^a \gamma_\mu b_{La}(x)\, \bar b_L^b
\gamma_\nu t_{Lb}(0)\rvac \nn\\
&=& \frac{d N_C\Gamma(2-d/2)}{4(4\pi)^{d/2}}\int_0^1 dx\, x \,
\left[\frac{2(p_\mu p_\nu - p^2 g_{\mu\nu})(1-x) + m_t^2 g_{\mu\nu}}{\left(m_t^2 x- p^2
  x(1-x)\right)^{(2-d/2)}}\right]\,.
\eea
Using
\be\label{eq:effLambda}
\frac{d\,\Gamma(2-d/2)(\Delta^2)^{(d/2-2)}}{4(4\pi)^{d/2}} =
\frac{2}{16\pi^2}\left[\epsilon^{-1} -\thalf\gamma + \thalf \ln 4\pi
    - \tfourth - \thalf\ln \Delta^2 + \CO(\epsilon)\right] \longrightarrow
  \frac{1}{16\pi^2} \ln (\Lambda^2/\Delta^2)\,, 
\ee
we get for the sum of the quark and technifermion loops:
\bea\label{eq:PiEW}
\Pi^{\pm}_{\mu\nu}(p) &=& \frac{N_C}{16\pi^2}\int_0^1 dx\, x
\left[2(p_\mu p_\nu - p^2 g_{\mu\nu})(1-x) + m_t^2 g_{\mu\nu}\right]\,
\ln\left(\frac{\Lambda^2}{m_t^2 x- p^2 x(1-x)}\right) \nn\\
&+& (N_C, m_t \to N_{TC}, m_U)\,.
\eea

The charged Goldstone-boson pole contribution to $\Pi^{\pm}_{\mu\nu}$ is
\bea\label{eq:PitbGB}
\Pi^{\pm}_{\mu\nu},_{GB}(p) &=&
\frac{i}{g_2^2}\left(\frac{g_2}{\sqrt{2}}\right)^2
\int d^4x d^4y\, e^{ip\cdot(x-y)}\lvac T\left[j_{L\mu}^{(1+i2)}(x)
\left(i \CL_{ETC}(0) + \cdots\right) j_{L\nu}^{(1-i2)}(0)\right]\rvac \nn\\
&=& \frac{2p_\mu p_\nu}{(16\pi^2)^2} \biggl\{\biggl[\int_0^1 dx\, x N_C m_t
    \ln\left(\frac{\Lambda^2}{m_t^2 x - p^2 x(1-x)}\right)\biggr]^2 \left(-4
    \Gamma_{0^-}^{t\bar b}(p)\right)\nn\\
&+& (t\bar b \leftrightarrow U\bar D) + (U\bar d \to U\bar D) \,\,{\rm terms}
\biggr\}\\ 
&=& -\frac{p_\mu p_\nu}{16\pi^2 p^2}\int_0^1 dx\, x \left[N_C m_t^2
  \ln\left(\frac{\Lambda^2}{m_t^2 x-p^2 x(1-x)}\right) + (N_C, m_t \to
  N_{TC}, m_U)\right]\,. \nn
\eea
The factor of $-4\Gamma_{0^-}^{t\bar b}$ comes from the first term on the
right in Eq.~(\ref{eq:tbGBa}), which indicates that $G_1 +\cdots$ sums to
this. These GB-pole terms combine with the $m^2 g_{\mu\nu}$-terms in
Eq.~(\ref{eq:PiEW}) to make a transverse massless-GB pole term. Then, with
Eq.~(\ref{eq:Wprop}), $g_{2W}^{-2}(p^2)$ and $f_W^2(p^2)$ are easily read
off, and are given in Eqs.~(\ref{eq:gtwoW},\ref{eq:fW}).

The calculations for the neutral EW propagator matrix are similar, if more
tedious. The important thing there is to arrange the terms so that there is a
massless photon pole.

\section*{Acknowledgments}

I have benefited from many conversations with Chris Hill and others,
including T.~Appelquist, G.~Burdman, R.~S.~Chivukula, E.~Eichten, A.~Martin,
E.~Pilon and B.~Zhou. I gratefully acknowledge the support of this project by
the Labex ENIGMASS during 2013 and 2014. I also thank Laboratoire
d'Annecy-le-Vieux de Physique Th\'eorique (LAPTh) for its hospitality and the
CERN Theory Group for support and hospitality during this research. My
research is supported in part by the U.S.~Department of Energy under Grant
No,~DE-SC0010106.


\bibliography{TCETC_Higgs_1}

\providecommand{\href}[2]{#2}\begingroup\raggedright\begin{thebibliography}{10}

\bibitem{Wilson:1970ag}
K.~G. Wilson, ``{The Renormalization Group and Strong Interactions},'' {\em
  Phys.Rev.} {\bf D3} (1971) 1818.

\bibitem{'tHooft:1979bh}
G.~'t~Hooft, ``{Naturalness, chiral symmetry, and spontaneous chiral symmetry
  breaking},'' {\em NATO Adv.Study Inst.Ser.B Phys.} {\bf 59} (1980) 135.

\bibitem{Weinberg:1979bn}
S.~Weinberg, ``Implications of Dynamical Symmetry Breaking: an addendum,'' {\em
  Phys. Rev.} {\bf D19} (1979) 1277--1280.

\bibitem{Susskind:1978ms}
L.~Susskind, ``Dynamics of Spontaneous Symmetry Breaking in the Weinberg-Salam
  Theory,'' {\em Phys. Rev.} {\bf D20} (1979) 2619--2625.

\bibitem{Englert:1964et}
F.~Englert and R.~Brout, ``{Broken Symmetry and the Mass of Gauge Vector
  Mesons},'' {\em Phys.Rev.Lett.} {\bf 13} (1964) 321--323.

\bibitem{Higgs:1964pj}
P.~W. Higgs, ``{Broken Symmetries and the Masses of Gauge Bosons},'' {\em
  Phys.Rev.Lett.} {\bf 13} (1964) 508--509.

\bibitem{Guralnik:1964eu}
G.~Guralnik, C.~Hagen, and T.~Kibble, ``{Global Conservation Laws and Massless
  Particles},'' {\em Phys.Rev.Lett.} {\bf 13} (1964) 585--587.

\bibitem{Weinberg:1967tq}
S.~Weinberg, ``{A Model of Leptons},'' {\em Phys.Rev.Lett.} {\bf 19} (1967)
  1264--1266.

\bibitem{Salam:1968rm}
A.~Salam, ``{Weak and Electromagnetic Interactions},'' {\em Conf.Proc.} {\bf
  C680519} (1968) 367--377.

\bibitem{Aad:2012tfa}
{\bf ATLAS Collaboration} Collaboration, G.~Aad {\em et.~al.}, ``{Observation
  of a new particle in the search for the Standard Model Higgs boson with the
  ATLAS detector at the LHC},'' {\em Phys.Lett.} {\bf B716} (2012) 1--29,
  \href{http://xxx.lanl.gov/abs/1207.7214}{ 1207.7214}.

\bibitem{Chatrchyan:2012ufa}
{\bf CMS Collaboration} Collaboration, S.~Chatrchyan {\em et.~al.},
  ``{Observation of a new boson at a mass of 125 GeV with the CMS experiment at
  the LHC},'' {\em Phys.Lett.} {\bf B716} (2012) 30--61,
  \href{http://xxx.lanl.gov/abs/1207.7235}{ 1207.7235}.

\bibitem{Giudice:2013yca}
G.~F. Giudice, ``{Naturalness after LHC8},''
  \href{http://xxx.lanl.gov/abs/1307.7879}{ 1307.7879}.

\bibitem{Bellazzini:2014yua}
B.~Bellazzini, C.~Cs\'aki, and J.~Serra, ``{Composite Higgses},''
  \href{http://xxx.lanl.gov/abs/1401.2457}{ 1401.2457}.

\bibitem{Bardeen:1989ds}
W.~A. Bardeen, C.~T. Hill, and M.~Lindner, ``{Minimal Dynamical Symmetry
  Breaking of the Standard Model},'' {\em Phys.Rev.} {\bf D41} (1990) 1647.

\bibitem{Appelquist:1988as}
T.~Appelquist, M.~Einhorn, T.~Takeuchi, and L.~Wijewardhana, ``{Higher Mass
  Scales and Mass Hierarchies},'' {\em Phys.Lett.} {\bf B220} (1989) 223--228.

\bibitem{Takeuchi:1989qa}
T.~Takeuchi, ``{Analytical and Numerical Study of the Schwinger-dyson Equation
  With Four Fermion Coupling},'' {\em Phys.Rev.} {\bf D40} (1989) 2697.

\bibitem{Chivukula:1990bc}
R.~S. Chivukula, A.~G. Cohen, and K.~D. Lane, ``{Aspects of Dynamical
  Electroweak Symmetry Breaking},'' {\em Nucl.Phys.} {\bf B343} (1990)
  554--570.

\bibitem{Lane:1991qh}
K.~D. Lane and M.~Ramana, ``{Walking technicolor signatures at hadron
  colliders},'' {\em Phys.Rev.} {\bf D44} (1991) 2678--2700.

\bibitem{Appelquist:1997fp}
T.~Appelquist, J.~Terning, and L.~Wijewardhana, ``{Postmodern technicolor},''
  {\em Phys.Rev.Lett.} {\bf 79} (1997) 2767--2770,
  \href{http://xxx.lanl.gov/abs/hep-ph/9706238}{ hep-ph/9706238}.

\bibitem{Holdom:1981rm}
B.~Holdom, ``Raising the Sideways Scale,'' {\em Phys. Rev.} {\bf D24} (1981)
  1441.

\bibitem{Appelquist:1986an}
T.~W. Appelquist, D.~Karabali, and L.~C.~R. Wijewardhana, ``Chiral Hierarchies
  and the Flavor Changing Neutral Current Problem in Technicolor,'' {\em Phys.
  Rev. Lett.} {\bf 57} (1986) 957.

\bibitem{Yamawaki:1986zg}
K.~Yamawaki, M.~Bando, and K.-i. Matumoto, ``Scale Invariant Technicolor Model
  and a Technidilaton,'' {\em Phys. Rev. Lett.} {\bf 56} (1986) 1335.

\bibitem{Akiba:1986rr}
T.~Akiba and T.~Yanagida, ``Hierarchic Chiral Condensate,'' {\em Phys. Lett.}
  {\bf B169} (1986) 432.

\bibitem{Eichten:1980ah}
E.~Eichten and K.~Lane, ``Dynamical Breaking of Weak Interaction Symmetries,''
  {\em Phys. Lett.} {\bf B90} (1980) 125--130.

\bibitem{Cohen:1988sq}
A.~G. Cohen and H.~Georgi, ``{Walking Beyond the Rainbow},'' {\em Nucl.Phys.}
  {\bf B314} (1989) 7.

\bibitem{Nambu:1988mr}
Y.~Nambu, ``{QUASISUPERSYMMETRY, BOOTSTRAP SYMMETRY BREAKING AND FERMION
  MASSES},''.

\bibitem{Miransky:1988xi}
V.~Miransky, M.~Tanabashi, and K.~Yamawaki, ``{Dynamical Electroweak Symmetry
  Breaking with Large Anomalous Dimension and t Quark Condensate},'' {\em
  Phys.Lett.} {\bf B221} (1989) 177.

\bibitem{Miransky:1989ds}
V.~Miransky, M.~Tanabashi, and K.~Yamawaki, ``{Is the t Quark Responsible for
  the Mass of W and Z Bosons?},'' {\em Mod.Phys.Lett.} {\bf A4} (1989) 1043.

\bibitem{Dobrescu:1997nm}
B.~A. Dobrescu and C.~T. Hill, ``{Electroweak symmetry breaking via top
  condensation seesaw},'' {\em Phys.Rev.Lett.} {\bf 81} (1998) 2634--2637,
  \href{http://xxx.lanl.gov/abs/hep-ph/9712319}{ hep-ph/9712319}.

\bibitem{Chivukula:1998wd}
R.~S. Chivukula, B.~A. Dobrescu, H.~Georgi, and C.~T. Hill, ``{Top quark seesaw
  theory of electroweak symmetry breaking},'' {\em Phys.Rev.} {\bf D59} (1999)
  075003, \href{http://xxx.lanl.gov/abs/hep-ph/9809470}{ hep-ph/9809470}.

\bibitem{Fukano:2012qx}
H.~S. Fukano and K.~Tuominen, ``{A hybrid 4$^{\textrm{th}}$ generation:
  Technicolor with top-seesaw},'' {\em Phys.Rev.} {\bf D85} (2012) 095025,
  \href{http://xxx.lanl.gov/abs/1202.6296}{ 1202.6296}.

\bibitem{Fukano:2013kia}
H.~S. Fukano and K.~Tuominen, ``{126 GeV Higgs boson in the top-seesaw
  model},'' {\em JHEP} {\bf 1309} (2013) 021,
  \href{http://xxx.lanl.gov/abs/1306.0205}{ 1306.0205}.

\bibitem{Bar-Shalom:2013hda}
S.~Bar-Shalom, ``{Dynamical Origin for the 125 GeV Higgs; a Hybrid setup},''
  \href{http://xxx.lanl.gov/abs/1310.2942}{ 1310.2942}.

\bibitem{Geller:2013dla}
M.~Geller, S.~Bar-Shalom, and A.~Soni, ``{Hybrid dynamical electroweak symmetry
  breaking with heavy quarks and the 125 GeV Higgs},''
  \href{http://xxx.lanl.gov/abs/1302.2915}{ 1302.2915}.

\bibitem{DiChiara:2014gsa}
S.~Di~Chiara, R.~Foadi, and K.~Tuominen, ``{125 GeV Higgs from a
  chiral-techniquark model},'' \href{http://xxx.lanl.gov/abs/1405.7154}{
  1405.7154}.

\bibitem{Peskin:1990zt}
M.~E. Peskin and T.~Takeuchi, ``A new constraint on a strongly interacting
  Higgs sector,'' {\em Phys. Rev. Lett.} {\bf 65} (1990) 964--967.

\bibitem{Golden:1990ig}
M.~Golden and L.~Randall, ``Radiative corrections to electroweak parameters in
  technicolor theories,'' {\em Nucl. Phys.} {\bf B361} (1991) 3--23.

\bibitem{Holdom:1990tc}
B.~Holdom and J.~Terning, ``Large corrections to electroweak parameters in
  technicolor theories,'' {\em Phys. Lett.} {\bf B247} (1990) 88--92.

\bibitem{Altarelli:1991fk}
G.~Altarelli, R.~Barbieri, and S.~Jadach, ``Toward a model independent analysis
  of electroweak data,'' {\em Nucl. Phys.} {\bf B369} (1992) 3--32.

\bibitem{Lane:1994pg}
K.~D. Lane, ``{Technicolor and precision tests of the electroweak
  interactions},'' \href{http://xxx.lanl.gov/abs/hep-ph/9409304}{
  hep-ph/9409304}.

\bibitem{Lane:2002wv}
K.~Lane, ``{Two lectures on technicolor},''
  \href{http://xxx.lanl.gov/abs/hep-ph/0202255}{ hep-ph/0202255}.

\bibitem{Appelquist:2010xv}
{\bf LSD Collaboration} Collaboration, T.~Appelquist {\em et.~al.}, ``{Parity
  Doubling and the S Parameter Below the Conformal Window},'' {\em
  Phys.Rev.Lett.} {\bf 106} (2011) 231601,
  \href{http://xxx.lanl.gov/abs/1009.5967}{ 1009.5967}.

\bibitem{Schaich:2011qz}
{\bf LSD Collaboration} Collaboration, D.~Schaich, ``{S parameter and parity
  doubling below the conformal window},'' {\em PoS} {\bf LATTICE2011} (2011)
  087, \href{http://xxx.lanl.gov/abs/1111.4993}{ 1111.4993}.

\bibitem{Appelquist:2013sia}
T.~Appelquist, R.~Brower, S.~Catterall, G.~Fleming, J.~Giedt, {\em et.~al.},
  ``{Lattice Gauge Theories at the Energy Frontier},''
  \href{http://xxx.lanl.gov/abs/1309.1206}{ 1309.1206}.

\bibitem{Beringer:1900zz}
{\bf Particle Data Group} Collaboration, J.~Beringer {\em et.~al.}, ``{Review
  of Particle Physics (RPP)},'' {\em Phys.Rev.} {\bf D86} (2012) 010001.

\bibitem{Lane:2002sm}
K.~Lane and S.~Mrenna, ``{The Collider phenomenology of technihadrons in the
  technicolor straw man model},'' {\em Phys.Rev.} {\bf D67} (2003) 115011,
  \href{http://xxx.lanl.gov/abs/hep-ph/0210299}{ hep-ph/0210299}.

\bibitem{Brooijmans:2008se}
G.~H. Brooijmans, A.~Delgado, B.~A. Dobrescu, C.~Grojean, M.~Narain, {\em
  et.~al.}, ``{New Physics at the LHC: A Les Houches Report. Physics at TeV
  Colliders 2007 -- New Physics Working Group},''
  \href{http://xxx.lanl.gov/abs/0802.3715}{ 0802.3715}.

\end{thebibliography}\endgroup
\bibliographystyle{utcaps}
\end{document}